\documentclass[final,1p,times]{elsarticle}
\usepackage{ecrc}

\volume{00}
\firstpage{1}
\journal{Nuclear Physics A}
\jid{nps}

\usepackage{graphicx}
\usepackage{dcolumn}
\usepackage{amsfonts}
\usepackage{amssymb,amsmath}

\biboptions{sort&compress} 

\begin{document}

\begin{frontmatter}

\title{Time-dependent HF approach to SHE dynamics}
\author[label1]{A.S. Umar}
\ead{umar@compsci.cas.vanderbilt.edu}

\author[label1]{V.E. Oberacker}
\address[label1]{Department of Physics and Astronomy, Vanderbilt University, Nashville, TN 37235, USA}

\begin{abstract} We employ the time-dependent Hartree-Fock (TDHF) method to study various
aspects of the reactions utilized in searches for superheavy elements. These include capture
cross-sections, quasifission, prediction of $P_{\mathrm{CN}}$, and other interesting dynamical quantities.
We show that the microscopic TDHF approach provides an important tool to
shed some light on the nuclear dynamics leading to the formation of superheavy elements.
\end{abstract}


\end{frontmatter}

\section{Introduction}

The search for new elements is one of the most novel and challenging research
areas of nuclear physics~\cite{armbruster1985,hofmann1998,hofmann2000,oganessian2007}.
The discovery of a region of the nuclear chart that can sustain the so
called \textit{superheavy elements} (SHE) has lead to intense experimental activity
resulting in the discovery of elements with atomic numbers as large as
$Z=117$~\cite{oganessian2010,oganessian2012,hinde2014}.
The theoretically predicted \textit{island of stability} is the result of new
proton and neutron shell-closures, whose location is not precisely
known~\cite{staszczak2013,cwiok2005,burvenich2004,moller2009,pei2009a,abusara2012}.
The experiments to discover these new elements are notoriously difficult, with
production cross-sections in pico-barns.
Of primary importance for the experimental investigations appear to be the choice
of target-projectile combinations that have the highest probability for forming
a compound nucleus that results in the production of the desired element.
Experimentally, two approaches have been used for the synthesis of these elements,
one utilizing doubly-magic $^{208}$Pb targets or $^{209}$Bi (cold-fusion)~\cite{hofmann2000,hofmann2002,naik2007}, the other
utilizing deformed actinide targets with neutron-rich projectiles (hot-fusion), such as $^{48}$Ca
~\cite{oganessian2004a,oganessian2007,hofmann2007}.
While both methods have been successful in synthesizing new elements the evaporation
residue cross-sections for hot-fusion were found to be several
orders of magnitude larger than those for cold fusion.
To pinpoint the root of this difference it is important to understand the details
of the reaction dynamics of these systems.
For light and medium mass systems the capture cross-section may be considered
to be the same as that for complete fusion, whereas for heavy systems leading to superheavy
formations the evaporation residue cross-section is dramatically reduced due to the
quasifission (QF) and fusion-fission
processes~\cite{hinde1995,hinde2002,itkis2011,knyazheva2007,shen1987,nishio2012,williams2013}
thus making the capture cross-section to
be essentially the sum of these two cross-sections, with QF occurring at a much shorter time-scale..
Consequently, quasifission is the primary reaction mechanism that limits the formation of
superheavy nuclei.
Various theoretical models have been developed to study the quasifission
process~\cite{adamian2003,adamian2009,nasirov2009a,zagrebaev2007,fazio2005,aritomo2009,zagrebaev2013}.

In many branches of science, highly complex many-body systems are often
described in macroscopic terms, which is particularly true
in the case of non-relativistic heavy-ion
collisions.
For example, the time evolution of the collective nuclear surface variables
$\alpha_{\ell m}(t)$ and the
corresponding geometrical nuclear shape $R(\theta, \phi, t)$ provides a very useful
set of parameters
to help organize experimental data. Using this approach
numerous evolutionary models
have been developed to explain particular aspects of
experimental data.
These methods provide a useful and productive means for
quantifying multitudinous reaction data.
In practice, they require a quantitative
understanding of the data as well as a clear physical picture of the important
aspects of the reaction dynamics.
The depiction of the collision must
be given at the onset, including the choice of coordinates which govern
the evolution of the reaction. Guessing the correct degrees of
freedom is extremely hard, without a full understanding of the dynamics, and
can easily lead to misbegotten results. More importantly, it is most often
not possible to connect these
macroscopic classical parameters, describing nuclear
matter under extreme excitation and rearrangement,
with the more fundamental properties of the nuclear force.
Such difficulties can
only be overcome with a fully microscopic theory of the collision dynamics.

The theoretical formalism for the
microscopic description of complex many-body quantum systems
and the understanding of the nuclear interactions that result in
self-bound, composite nuclei possessing the observed properties
are the underlying challenges for studying low energy nuclear physics.
The Hartree-Fock approximation
and its time-dependent generalization, the time-dependent Hartree-Fock
theory, has provided a possible means to study the diverse phenomena
observed in low energy nuclear physics~\cite{negele1982,simenel2012}.
In general TDHF theory provides a useful foundation for a
fully microscopic many-body theory of large amplitude collective
motion including collective surface vibrations and giant
resonances~\cite{umar2005a,maruhn2005,nakatsukasa2005,simenel2003,reinhard2006,reinhard2007,suckling2010,pei2005,xu2005},
nuclear reactions in the
vicinity of the Coulomb barrier, such as fusion~\cite{umar2010a,washiyama2008,keser2012,simenel2008,umar2014a},
deep-inelastic reactions and transfer~\cite{simenel2010,simenel2011,umar2008a,sekizawa2013,scamps2013,simenel2013b},
and dynamics of fission fragments~\cite{simenel2014a}.
As a result of theoretical approximations (single Slater determinant), TDHF does not
describe individual reaction channels; rather, it describes the
time-evolution of the dominant reaction channel.
In other words TDHF is a deterministic theory. To obtain multiple reaction channels or widths of
observables one must go beyond TDHF~\cite{tohyama2002a,tohyama2001,lacroix2014}.
In connection with
superheavy element formation, the theory predicts
best the cross-section for a particular process which dominates the
reaction mechanism. This is certainly the case for studying capture
cross-sections and quasifission.

In recent years has it become numerically feasible to perform TDHF calculations on a
3D Cartesian grid without any symmetry restrictions
and with much more accurate numerical methods~\cite{umar1991a,umar2006c,maruhn2014,simenel2012}.
In addition, the quality of effective interactions has been substantially
improved~\cite{chabanat1998a,kluepfel2009,kortelainen2010}.
While ordinary TDHF calculations can be used for fusion or capture at energies above the barrier
they cannot be used directly at sub-barrier energies.
During the past several years, a novel approach based on TDHF called the density constrained
time-dependent Hartree-Fock (DC-TDHF) method was developed to compute
heavy-ion potentials~\cite{umar2006b} and excitation energies~\cite{umar2009a} directly from TDHF
time-evolution. This method was applied
to calculate fusion and capture cross sections above and below the barrier,
ranging from light and medium mass systems~\cite{umar2006d,umar2008b,umar2009b,keser2012,simenel2013a,umar2012a}
to hot and cold fusion reactions leading to superheavy element $Z=112$~\cite{umar2010a}.
In all cases a
good agreement between the measured fusion cross sections and the DC-TDHF results was found.
This is rather remarkable given the fact that the only input in DC-TDHF is the
Skyrme effective N-N interaction, whose parameters are determined from static structure
information and there are no adjustable parameters.

Within the last few years the TDHF approach has been utilized for studying the dynamics of
quasifission~\cite{wakhle2014,oberacker2014}
and scission~\cite{simenel2014a}. Particularly, the study of quasifission is showing a great promise to provide
insight based on very favorable comparisons with experimental data. In this article we will
focus on the TDHF studies of capture cross-sections, quasifission observables, and related quantities.

\section{Theory}
We now give a brief outline of the TDHF method and some of the recent extensions used in the
calculations presented~\cite{negele1982,simenel2012}.
Given a many-body Hamiltonian $H$ containing two and three-body interactions
the time-dependent action $S$ can be constructed as
\begin{equation}
S=\int_{t_1}^{t_2}dt<\Phi(t)|H-i\hbar\partial_t|\Phi(t)>\;.
\end{equation}
Here, $\Phi$ denotes the time-dependent many-body
wavefunction $\Phi({\bf r_1,r_2,\ldots,r_{A}};t)$.
General variation of $S$ recovers the time-dependent Schr\"odinger
equation.
In TDHF approximation the many-body wavefunction is replaced by a single
Slater determinant and this form is preserved at all times.
The determinental form guarantees the antisymmetry required by the Pauli
principle for a system of fermions. In this limit, the
variation of the action yields the most probable time-dependent path
between points $t_1$ and $t_2$ in the multi-dimensional
space-time phase space
\begin{equation}
\delta S=0 \rightarrow \Phi(t)=\Phi_0(t)\;.
\label{variat}
\end{equation}
In practice $\Phi_0(t)$ is chosen to be a Slater determinant comprised of
single-particle states $\phi_{\lambda}(\mathbf{r},t)$ with
quantum numbers $\lambda$.
If the variation in Eq.(\ref{variat}) is performed with respect to
the single-particle states $\phi^{*}_{\lambda}$ we obtain a
set of coupled, nonlinear, self-consistent initial value equations
for the single-particle states
\begin{equation}
h\left( \left\{ \phi_{\mu} \right\} \right) \phi_{\lambda}=i\hbar
\dot{\phi_{\lambda}}
\;\;\;\;\;\;\;\;\;\lambda=1,...,N\;.
\label{tdhf0}
\end{equation}
These are the fully microscopic time-dependent Hartree-Fock equations
which preserve the major conservation laws such as the particle number,
total energy, total angular momentum, etc.
As we see from Eq.(\ref{tdhf0}), each single-particle state evolves in the
mean-field $h$ generated by the concerted action of all the other single-particle
states.

In TDHF, the initial nuclei are calculated by solving the static Hartree-Fock (HF)
equations using the damped-relaxation method~\cite{bottcher1989,strayer1990}.
The resulting Slater
determinants for each nucleus comprise the larger Slater determinant describing the colliding
system during the TDHF evolution.
Nuclei are assumed
to move on a pure Coulomb trajectory until the initial separation between the nuclear centers used
in TDHF evolution. Using the Coulomb trajectory we compute the relative kinetic energy at this
separation and the associated translational momenta for each nucleus. The nuclei are than boosted
by multiplying the HF states with a phase factor
\begin{equation}
\Phi _{j}\rightarrow \exp (\imath\mathbf{k}_{j}\cdot \mathbf{R})\Phi _{j}\;,
\end{equation}
where $\Phi _{j}$ is the HF state for nucleus $j$ and $\mathbf{R}$ is the corresponding
center of mass coordinate
\begin{equation}
\mathbf{R}=\frac{1}{A_{j}}\sum _{i=1}^{A_{j}}\mathbf{r}_{i}\;.
\end{equation}
The Galilean invariance of the TDHF equations assures the evolution of
the system without spreading and the conservation of the total energy for the system.
In TDHF, the many-body state remains a Slater determinant at all times.

\subsection{DC-TDHF method}

The concept of using density as a constraint for calculating collective states
from TDHF time-evolution was first introduced in~\cite{cusson1985}, and used
in calculating collective energy surfaces in connection with nuclear molecular
resonances in~\cite{umar1985}.

In this approach we assume that a collective state is characterized only by
density  $\rho$ and current $\mathbf{j}$. This state can be constructed
by solving the static Hartree-Fock equations
\begin{equation}
<\Phi_{\rho,\mathbf{j}}|a_h^{\dagger}a_p\hat{H}|\Phi_{\rho,\mathbf{j}}>=0\;,
\end{equation}
subject to constraints on
density and current
\begin{eqnarray*}
	<\Phi_{\rho,\mathbf{j}}|\hat{\rho}(\mathbf{r})|\Phi_{\rho,\mathbf{j}}>&=&\rho(\mathbf{r},t) \\
	<\Phi_{\rho,\mathbf{j}}|\hat{\jmath}(\mathbf{r})|\Phi_{\rho,\mathbf{j}}>&=&\mathbf{j}(\mathbf{r},t)\;.
\end{eqnarray*}
Choosing $\rho(\mathbf{r},t) $ and $\mathbf{j}(\mathbf{r},t)$ to be the instantaneous TDHF
density and current results in the lowest energy collective state corresponding to the
instantaneous TDHF state $|\Phi(t)>$, with the corresponding energy
\begin{equation}
E_{coll}(\rho(t),\mathbf{j}(t))=<\Phi_{\rho,\mathbf{j}}|\hat{H}|\Phi_{\rho,\mathbf{j}}>\;.
\end{equation}
This collective energy differs from the conserved TDHF energy only by the amount of
internal excitation present in the TDHF state, namely
\begin{equation}
E^{*}(t)=E_{TDHF} - E_{coll}(t)\;.
\end{equation}
However, in practical calculations the constraint on the current is difficult to implement
but we can define instead a static adiabatic collective state $|\Phi_{\rho}>$ subject to the
constraints
\begin{eqnarray*}
	<\Phi_{\rho}|\hat{\rho}(\mathbf{r})|\Phi_{\rho}>&=&\rho(\mathbf{r},t) \\
	<\Phi_{\rho}|\hat{\jmath}(\mathbf{r})|\Phi_{\rho}>&=&0\;.
\end{eqnarray*}
In terms of this state one can write the collective energy as
\begin{equation}
\label{eq:4}
E_{coll}=E_{kin}(\rho(t),\mathbf{j}(t))+E_{DC}(\rho(\mathbf{r},t))\;,
\end{equation}
where the density-constrained energy $E_{DC}$, and the collective kinetic
energy $E_{kin}$ are defined as
\begin{eqnarray*}
	E_{DC}&=&<\Phi_{\rho}|\hat{H}|\Phi_{\rho}> \\
	E_{kin}&\approx&\frac{m}{2}\sum_q\int d^{3}r\; \mathbf{j}^2_q(t)/\rho_q(t)\;,
\end{eqnarray*}
where the index $q$ is the isospin index for neutrons and protons ($q=n,p$).
From Eq.~\ref{eq:4} is is clear that the density-constrained energy
$E_{DC}$ plays the role of a collective potential. In fact this is
exactly the case except for the fact that it contains the binding
energies of the two colliding nuclei. One can thus define the ion-ion
potential as~\cite{umar2006b}
\begin{equation}
V=E_{\mathrm{DC}}(\rho(\mathbf{r},t))-E_{A_{1}}-E_{A_{2}}\;,
\end{equation}
where  $E_{A_{1}}$ and $E_{A_{2}}$ are the binding energies of two nuclei
obtained from a static Hartree-Fock calculation with the same effective
interaction. For describing a collision of two nuclei one can label the
above potential with ion-ion separation distance $R(t)$ obtained during the
TDHF time-evolution. This ion-ion potential $V(R)$ is asymptotically correct
since at large initial separations it exactly reproduces $V_{Coulomb}(R_{max})$.
In addition to the ion-ion potential it is also possible to obtain coordinate
dependent mass parameters. One can compute the ``effective mass'' $M(R)$
using the conservation of energy in a central collision
\begin{equation}
M(R)=\frac{2[E_{\mathrm{c.m.}}-V(R)]}{\dot{R}^{2}}\;,
\label{eq:mr}
\end{equation}
where the collective velocity $\dot{R}$ is directly obtained from the TDHF time evolution and the potential
$V(R)$ from the density constraint calculations.

In practice, the potential barrier penetrabilities $T_L$ at $E_{\mathrm{c.m.}}$ energies
below and above the barrier are obtained by numerical integration of the
Schr\"odinger equation for the relative coordinate $R$ using
the well-established {\it Incoming Wave Boundary Condition} (IWBC) method~\cite{hagino2007}.

\subsection{Skyrme interaction}

Almost all TDHF calculations have been done using the Skyrme energy density functional.
The Skyrme energy density functional contains terms
which depend on the nuclear density, $\rho$, kinetic-energy density, $\tau$,
spin density, $\mathbf{s}$, spin kinetic energy density, $\mathbf{T}$, and the full spin-current pseudotensor,
$\mathbf{J}$, as
\begin{equation}
E=\int d^{3}r\;{\cal H}(\rho ,\tau,\mathbf{j},\mathbf{{s}},\mathbf{T},\mathbf{J};\mathbf{{r}})\;.
\end{equation}
The time-odd terms ($\mathbf{j}$, $\mathbf{s}$, $\mathbf{T}$) vanish
for static calculations of even-even nuclei, while they are present for odd mass nuclei, in cranking calculations, as well
as in TDHF. The spin-current pseudotensor, $\mathbf{J}$, is time-even and does not vanish for
static calculations of even-even nuclei.
It has been shown~\cite{umar1986a,reinhard1988,umar1989,umar2006c,suckling2010,fracasso2012} that the presence of these
extra terms are necessary for preserving the Galilean
invariance and make an appreciable contribution to the dissipative properties of the collision.
Our TDHF program includes all of the appropriate combinations of time-odd terms in the Skyrme interaction.
In addition, commonly a pairing force is added to incorporate pairing interactions for nuclei. The
implementation of pairing for time-dependent collisions is currently an unresolved problem although
small amplitude implementations exist~\cite{tohyama2002b,ebata2010}.
However, for reactions with relatively high excitation this is not expected to be a problem.

\section{Capture cross-sections}

For the reactions of heavy systems the process of overcoming the interaction barrier
is commonly referred to as \textit{capture}. After capture a number of possibilities
exist~\cite{yanez2013}
\begin{eqnarray*}
	\sigma_{\mathrm{capture}} &=& \sigma_{\mathrm{fusion}} + \sigma_{\mathrm{quasifission}} + \sigma_{\mathrm{fast fission}} \\
	\sigma_{\mathrm{fusion}} &=& \sigma_{\mathrm{evaporation\;residue}} + \sigma_{\mathrm{fusion-fission}}
\end{eqnarray*}
Understanding each of these cross-sections is vital for selecting reaction partners that have
the highest probability for producing a superheavy element.
As we have discussed above one of the applications of the DC-TDHF method is the calculation of
microscopic potential barriers for reactions leading to superheavy formations. This allows the
calculation of capture-cross sections and the excitation energy of the system at the capture point.
In this section we briefly describe capture cross-section calculations and in the following
sections we will show results for quasifission and the possibility of calculating some of the
ingredients for the calculation of $P_{\mathrm{CN}}$.

\begin{figure}[!htb]
	\begin{center}
			\includegraphics*[scale=0.31]{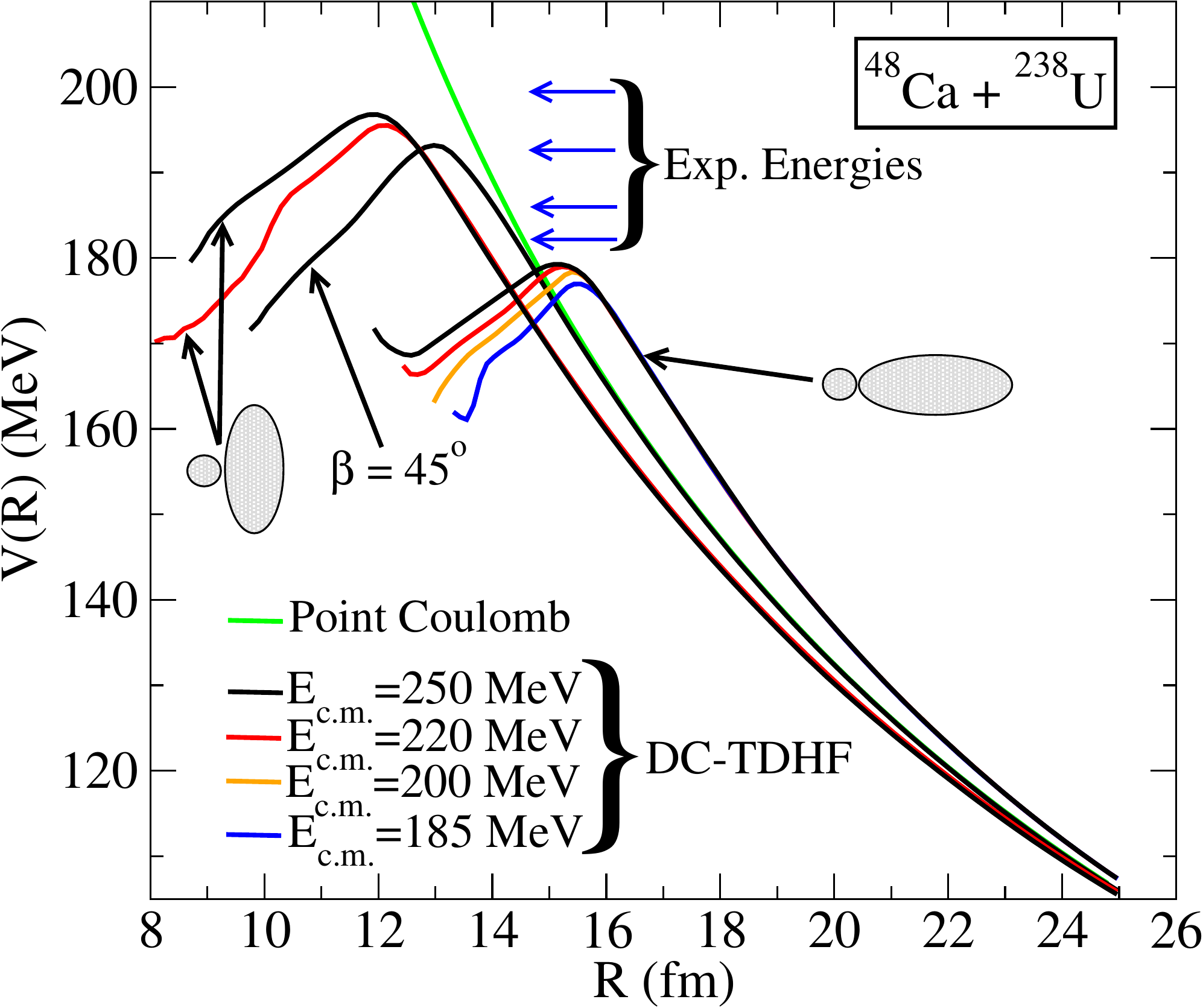}\hspace{0.2in}
			\includegraphics*[scale=0.31]{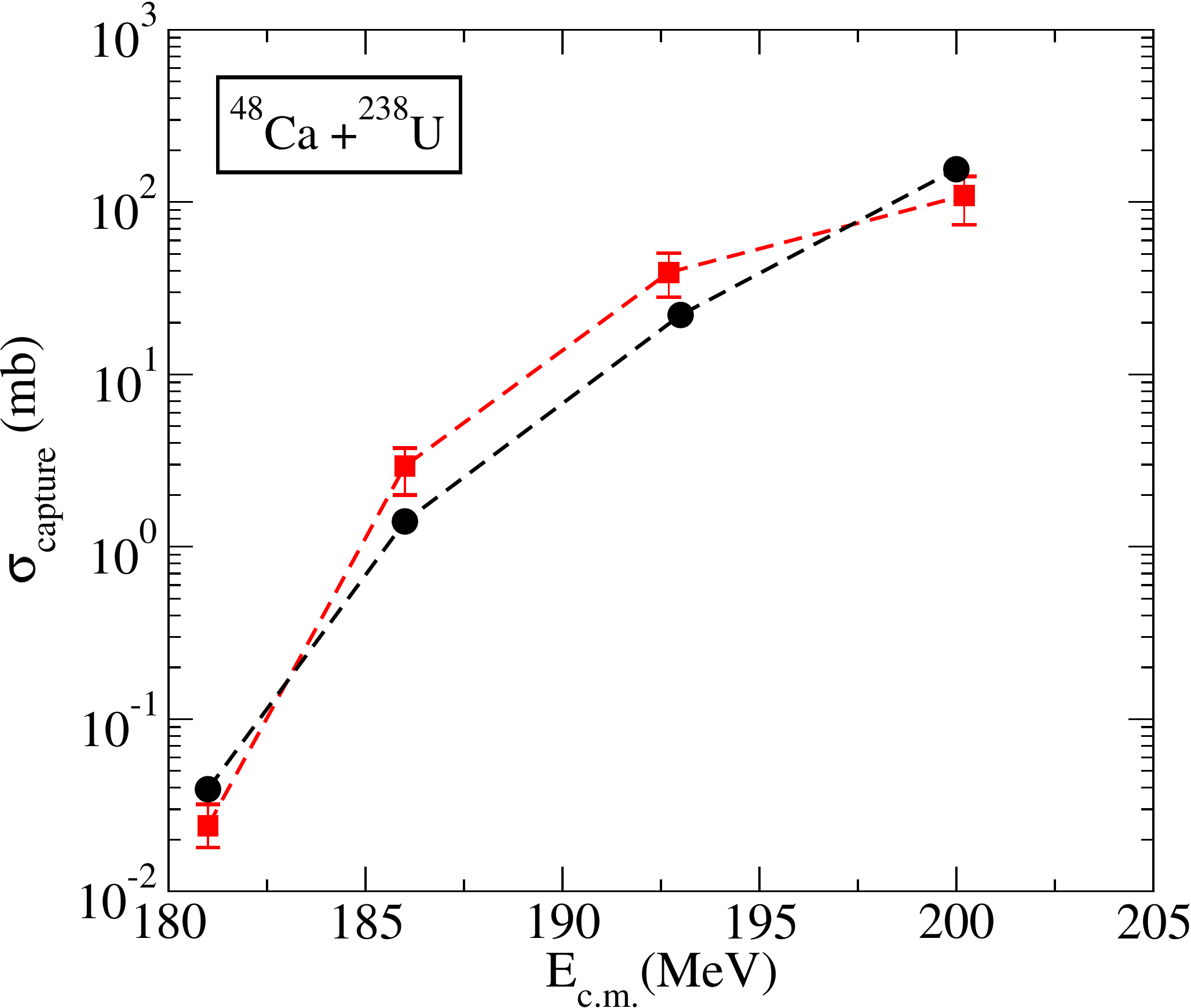}
	\end{center}
	\caption{\protect Left: Potential barriers, $V(R)$, obtained from
		DC-TDHF calculations~\cite{umar2010a} as a function of $E_\mathrm{c.m.}$ energy and
		orientation angle $\beta$ of the $^{238}$U nucleus. Also shown are the experimental c.m. energies.
		Right: Capture cross-sections as a function of $E_\mathrm{c.m.}$ energy (black circles).
		Also shown are the experimental cross-sections~\cite{itkis2002,itkis2004,oganessian2007} (red squares).}
	\label{fig:vrCa_U}
\end{figure}

In connection with superheavy element production, we have studied the hot fusion reaction
$^{48}$Ca+$^{238}$U and the cold fusion reaction $^{70}$Zn+$^{208}$Pb~\cite{umar2010a}.
Considering hot fusion,
$^{48}$Ca is a spherical nucleus whereas $^{238}$U has a large axial deformation.
The deformation of $^{238}$U strongly influences the interaction barrier for this
system. This is shown in the left panel of Fig.~\ref{fig:vrCa_U},
which shows the interaction barriers, $V(R)$,
calculated using the DC-TDHF method as a function of c.m. energy and for three different
orientations of the $^{238}$U nucleus. The alignment angle $\beta$ is the angle
between the symmetry axis of the $^{238}$U nucleus and the collision axis.
Also, shown in the left panel of Fig.~\ref{fig:vrCa_U} are the experimental energies~\cite{hofmann2007,oganessian2007} for this reaction.
We observe that all of the experimental energies are above the barriers obtained for the
polar alignment of the $^{238}$U nucleus.

The barriers for the polar orientation ($\beta=0^{o}$) of the
$^{238}$U nucleus are much lower and peak at larger ion-ion separation distance $R$.
On the other hand, the barriers for the equatorial orientation ($\beta=90^{o}$) are
higher and peak at smaller $R$ values.
We observe that at lower energies the polar orientation results in
sticking of the two nuclei, while the equatorial orientation results in a deep-inelastic
collision.
We have also calculated the excitation energy $E^{*}(R)$ as a function of c.m. energy
and orientation angle $\beta$ of the $^{238}$U nucleus.
The system is
excited much earlier during the collision process for the polar orientation
and has a higher excitation than the corresponding collision for the equatorial orientation.

To obtain the capture cross-section, we calculate potential barriers $V(R,\beta)$
for a set of initial orientations $\beta$ of the $^{238}$U nucleus. Then we determine
partial cross sections $\sigma(\beta)$ and perform an angle-average.
However, as a result of long-range Coulomb excitation, not all
initial orientation angles occur with the same
probability. Rather, the dominant excitation of the ground state
rotational band in deformed nuclei leads to a preferential alignment which
is calculated in a separate semiclassical Coulomb
excitation code~\cite{umar2006d}. This code is only used
to determine the alignment probability $\mathrm{d}P/(\mathrm{d}\beta\;\sin\beta)$ in Eq.~\ref{eq:scapture} for
the angle averaging of the cross-section
\begin{equation}
\label{eq:scapture}
\sigma_{\mathrm{capture}}(E_{\mathrm{c.m.}}) = \int_{0}^{\pi} \mathrm{d}\beta\;\sin\beta\; \frac{\mathrm{d}P}{\mathrm{d}\beta\sin\beta}\;
\sigma(E_{\mathrm{c.m.}},\beta)\;,
\end{equation}
and $\sigma(E_{\mathrm{c.m.}},\beta)$
is the capture cross-section associated with a particular alignment.
In the right panel of Fig.~\ref{fig:vrCa_U} we show our
results for the capture cross-sections which are in remarkably good agreement
with experimental data.

\section{Quasifission}
The feasibility of using TDHF for quasifission has only been
recognized recently~\cite{simenel2012,wakhle2014,oberacker2014}.
By virtue of long contact-times for quasifission and the energy
and impact parameter dependence these calculations require extremely long CPU times
and numerical accuracy~\cite{bottcher1989,reinhard1988,umar1989,umar1991a,umar2006c,maruhn2014}.

In the present TDHF calculations we use the Skyrme SLy4d interaction~\cite{kim1997}
including all of the relevant time-odd terms in the mean-field Hamiltonian.
First we generate very accurate static HF wave functions for the two nuclei on the
3D grid.
\begin{figure}[!htb]
	\centering
	\includegraphics*[width=11.9cm]{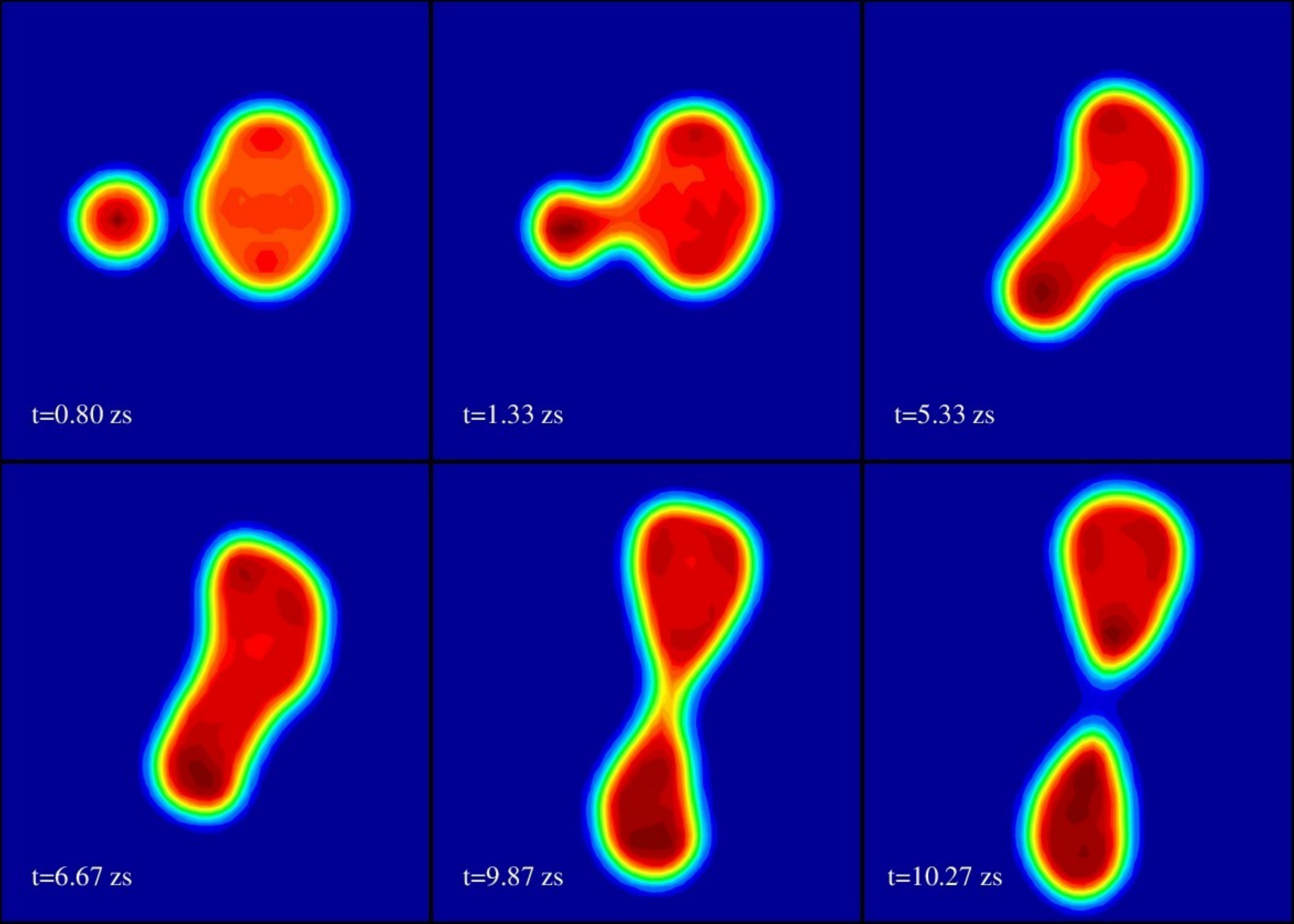}
	\caption{\protect Quasifission in the reaction $^{40}$Ca+$^{238}$U
		at $E_{\mathrm{c.m.}}=209$~MeV with impact parameter $b=1.103$~fm ($L=20$).
		Shown is a contour plot of the time evolution of the mass density. Time
		increases from left to right and top to bottom. The actual numerical box
		is larger than the ones shown in the figure.}
	\label{fig:dens1}
\end{figure}
The initial separation of the two nuclei is $30$~fm. In the second
step, we apply a boost operator to the single-particle wave functions. The time-propagation
is carried out using a Taylor series expansion (up to orders $10-12$) of the
unitary mean-field propagator, with a time step $\Delta t = 0.4$~fm/c.
In Fig.~\ref{fig:dens1} we show contour plots of the mass density
for the $^{40}$Ca+$^{238}$U reaction at $E_{\mathrm{c.m.}}=209$~MeV
as a function of time. The impact parameter $b=1.103$~fm corresponds to
an orbital angular momentum quantum number $L=20$. In this case, the 3D lattice
spans $(66 \times 56 \times 30)$~fm.
As the nuclei approach each other, a neck forms between the
two fragments which grows in size as the system begins to rotate. Due to the centrifugal forces
the dinuclear system elongates and forms a very long neck which eventually
ruptures leading to two separated fragments.
The $^{238}$U nucleus exhibits both quadrupole and hexadecupole
deformation; in the present study, its symmetry axis was
oriented initially at $90^{\circ}$ to the internuclear axis.
This orientation leads to the largest ``contact time''~\cite{simenel2012}
which is defined as the time interval between the time $t_1$
when the two nuclear surfaces first touch and
the time $t_2$ when the dinuclear system splits up again. In this case, we find
a contact time $\Delta t = 9.35$~zs (1~zs $ =10^{-21}$~s) and substantial mass transfer (66 nucleons
to the light fragment). The event has all the characteristics of quasifission.
The orientation of the $^{238}$U symmetry axis at $0^{\circ}$ to the internuclear axis also result in QF
but with much shorter contact-times and consequently with smaller mass transfer.
These contribute more to
the large asymmetric part of the mass distribution~\cite{wakhle2014} and will not be considered for the
purposes of this study.
\begin{figure}[!htb]
\centering
\includegraphics*[width=8.6cm]{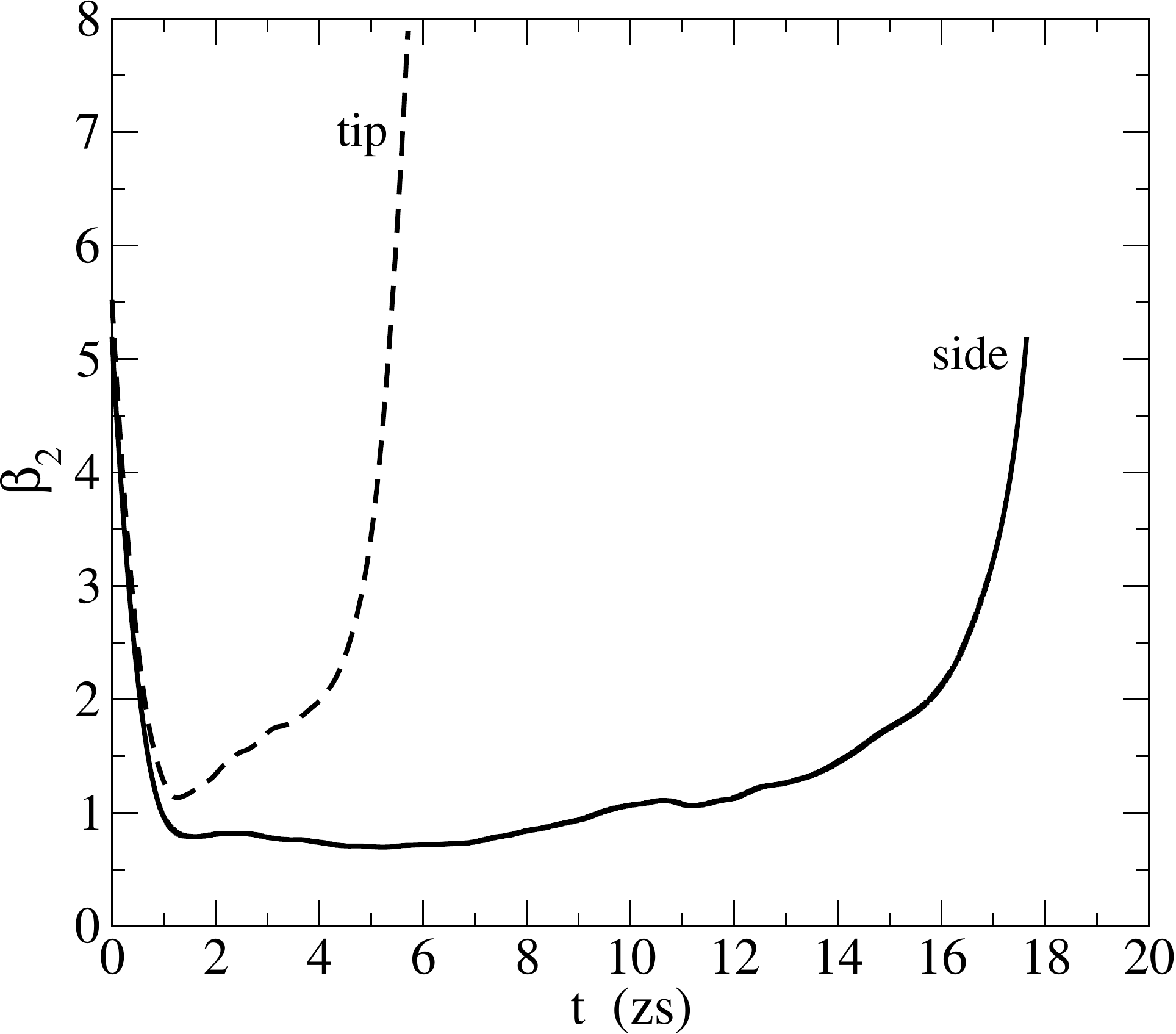}
\caption{\protect
TDHF results showing the time-dependence of the deformation $\beta_2$ for a central collision of $^{48}$Ca~+~$^{238}$U
at $E_\mathrm{c.m.}=203$~MeV. The two curves show the two orientations
of the $^{238}$U nucleus with respect to the collision axis.}
\label{fig:beta2}
\end{figure}

Another interesting observable is the time-evolution of the quasifissioning system.
In order to have the correct quadrupole moment for a changing nuclear density one has
to diagonalize the quadrupole tensor matrix
\begin{equation}
Q_{ij} = \int~d^3r\;\rho_{TDHF}(\mathbf{r},t) (3x_ix_j-r^2\delta_{ij})\;.
\end{equation}
The largest eigenvalue gives the quadrupole moment calculated along the
principal axis for the nucleus (after multiplying with $\sqrt{5/16\pi}$). The other
two eigenvalues allow the calculation of $Q_{22}$ as well.
From these one can construct the deformation parameter $\beta_2$
\begin{equation}
\beta_2=\frac{4\pi}{3}\frac{Q_{20}}{AR_0^2}\;,
\end{equation}
where $R_0=1.2A^{1/3}$.
The changing quadrupole moment during the collision, particularly towards the last
stages of the quasifission process shows not only the elongation of the nucleus but
the rate of change also shows the velocity by which the quasifission event is
taking place.
In Fig.~\ref{fig:beta2} we plot the TDHF results showing the time-dependence of the deformation
$\beta_2$ for a central collision of $^{48}$Ca~+~$^{238}$U
at $E_\mathrm{c.m.}=203$~MeV. The two curves show the two orientations
of the $^{238}$U nucleus with respect to the collision axis.
One clearly observes that the neutron-rich system stays at a compact shape much longer
than the neutron-poor system and the actual quasifission event happens relatively quickly.

\begin{figure}[!htb]
	\centering
	\includegraphics*[width=8.6cm]{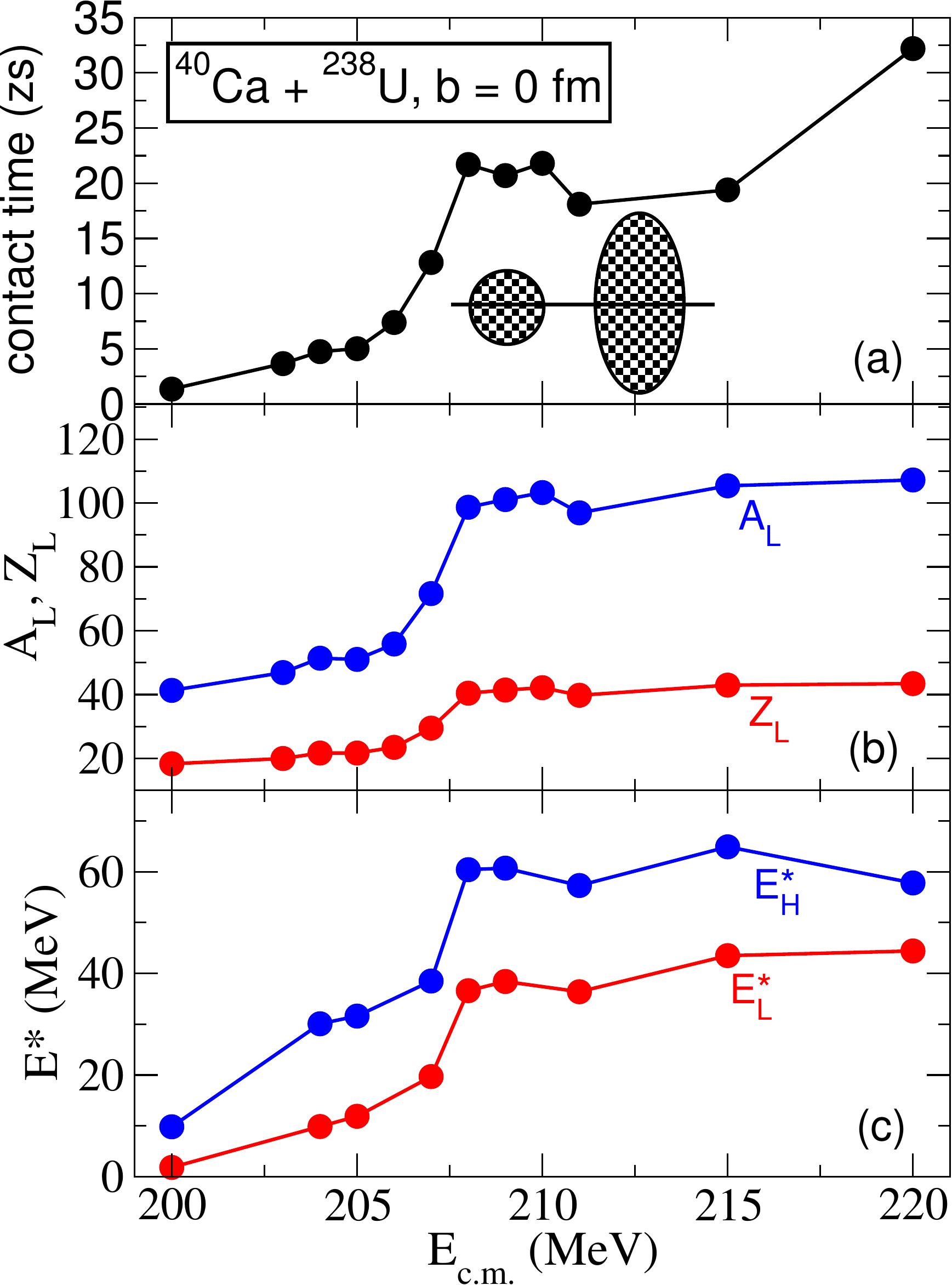}
	\caption{\protect Several observables as a function of
		center-of-mass energy for a central collision of $^{40}$Ca+$^{238}$U.
		(a) contact time, (b) mass and charge of the light fragment, and
		(c) excitation energy of the light and heavy fragments.
		Quasifission dominates in the energy region $E_{\mathrm{c.m.}}= 208-220$~MeV.}
	\label{fig:40Ca_stack_1}
\end{figure}
Next we consider central collisions of $^{40}$Ca+$^{238}$U.
In the energy interval $E_{\mathrm{c.m.}}= 200-220$~MeV we always observe two fragments in
the exit channel, i.e. these events are either a deep-inelastic reaction or
quasifission. In Fig.~\ref{fig:40Ca_stack_1}a we display the contact time as a function of the
center-of-mass energy.  We observe that in the energy interval
$E_{\mathrm{c.m.}}=200-205$~MeV the contact time increases slowly with increasing energy. Between
$E_{\mathrm{c.m.}}= 205-208$~MeV there is a steep increase in the contact time.
In the energy range $E_{\mathrm{c.m.}}=208-220$~MeV the contact time varies between
(21.7$-$32.2)~zs which is (16$-$24) times larger than
the contact time observed at $E_{\mathrm{c.m.}}= 200$~MeV.
The contact times observed in our TDHF calculations are of similar magnitude as
those obtained by Simenel (see Fig. 38 of Ref.~\cite{simenel2012} and
Fig. 8 of Ref.~\cite{simenel2012c}).
\begin{figure}[!htb]
	\centering
	\includegraphics*[width=8.6cm]{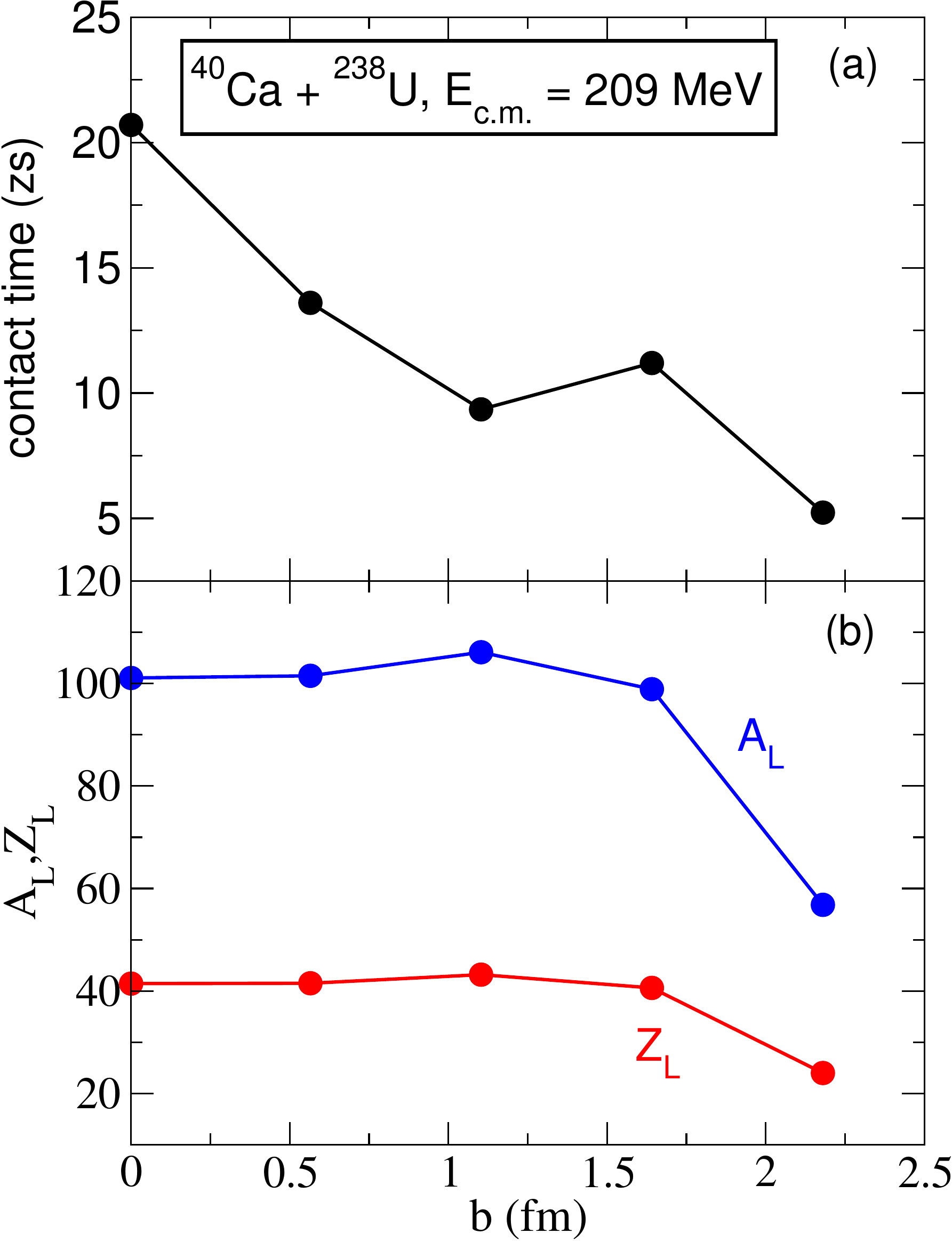}
	\caption{\protect (a) contact time and (b) mass and
		charge of the light fragment as a function of impact parameter.}
	\label{fig:40Ca_stack_2}
\end{figure}

In Fig.~\ref{fig:40Ca_stack_1}b we show the corresponding masses $A_\mathrm{L}$ and charges $Z_\mathrm{L}$ of
the light fragment. A comparison with Fig.~\ref{fig:40Ca_stack_1}a reveals that mass
and charge transfer are proportional to the contact time. For example, at
$E_{\mathrm{c.m.}}=200$~MeV there is very little mass transfer ($A_\mathrm{L}=41.4$) and
some charge pickup ($Z_\mathrm{L}=18.4$). At $E_{\mathrm{c.m.}}= 208$~MeV, the dramatic increase
in contact time results in a large amount of both mass and charge transfer, $A_\mathrm{L}=98.7$
and $Z_\mathrm{L}=40.5$. At $E_{\mathrm{c.m.}}= 220$~MeV we find a light fragment mass $A_\mathrm{L}=107.2$
and charge $Z_\mathrm{L}=43.4$. Based on these results, we conclude
that the energy region $E_{\mathrm{c.m.}}= 200-207$~MeV is likely dominated by
deep-inelastic reactions while the energy region $E_{\mathrm{c.m.}}= 208-220$~MeV
is dominated by quasifission. TDHF calculations carried out at higher energy,
$E_{\mathrm{c.m.}}= 223-225$~MeV show {\it one} fragment in the exit channel; this is
the fusion region. Naturally, quasifission is still possible at higher energies,
but only for a certain range of non-zero impact parameters.
\begin{figure}[!htb]
	\centering
	\includegraphics*[width=8.6cm]{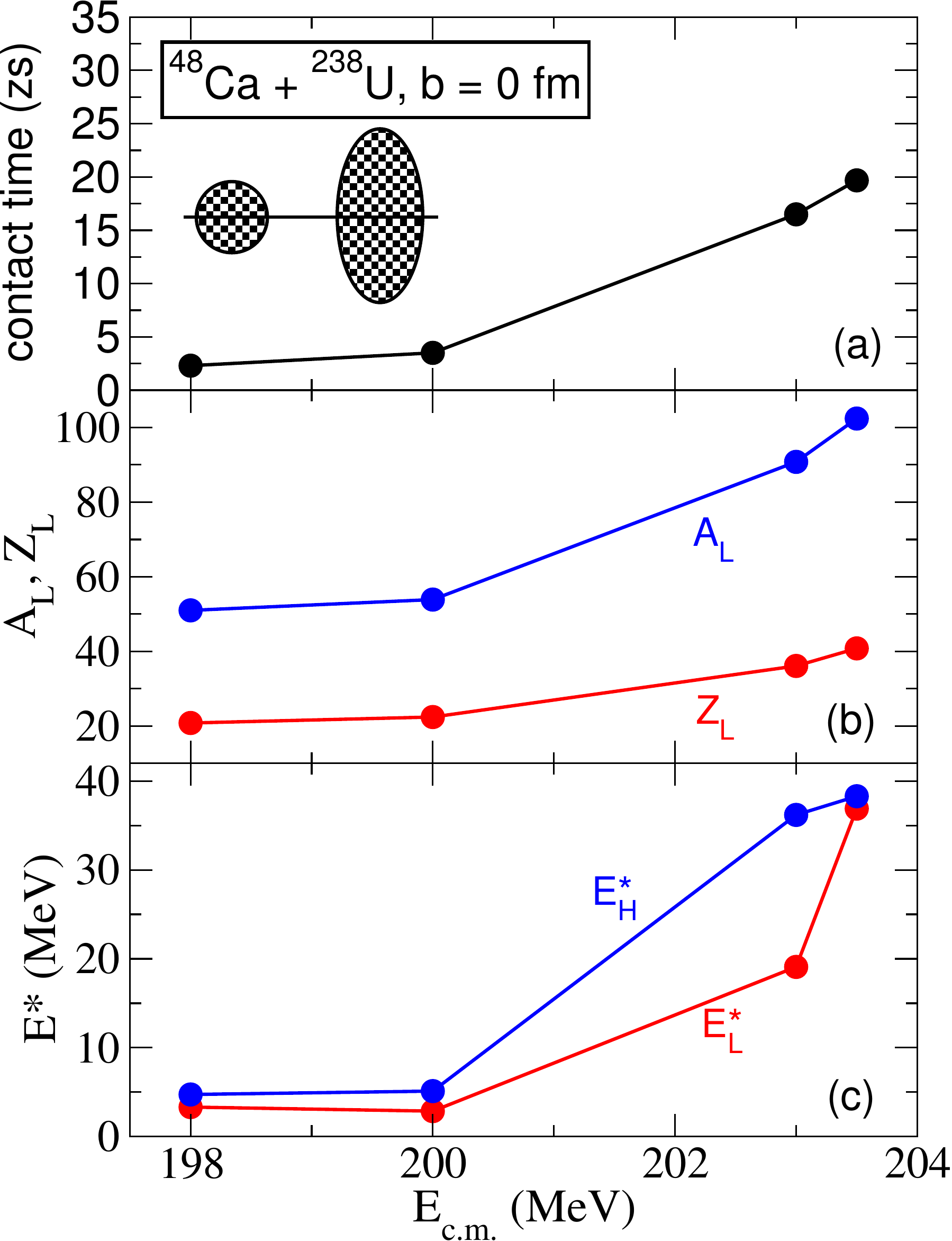}
	\caption{\protect Several observables as a function of
		center-of-mass energy for a central collision of $^{48}$Ca+$^{238}$U.
		(a) contact time, (b) mass and charge of the light fragment, and
		(c) excitation energy of the light and heavy fragments.
		Quasifission dominates in a very narrow energy window, $E_{\mathrm{c.m.}}=202-203.5$~MeV.}
	\label{fig:48Ca_stack_1}
\end{figure}
Recently, we have developed an extension to TDHF theory via the use
of a density constraint to calculate fragment excitation energies directly from the
TDHF time evolution~\cite{umar2009a}.
In Fig.~\ref{fig:40Ca_stack_1}c we show the excitation energies of the light and heavy fragments
as a function of the center-of-mass energy. For $^{40}$Ca+$^{238}$U, we find excitation energies
in the QF region to be as high as 60~MeV for the heavy fragment and
40~MeV for the light fragment.

Figure~\ref{fig:40Ca_stack_2}a shows the impact parameter dependence of the
contact time, and Fig.~\ref{fig:40Ca_stack_2}b exhibits the corresponding
masses $A_\mathrm{L}$ and charges $Z_\mathrm{L}$ of the light fragment for $^{40}$Ca+$^{238}$U
at a fixed  center-of-mass energy $E_{\mathrm{c.m.}}=209$ MeV.
We observe that the contact time decreases from its maximum around 21~zs for the central collision
to about 5~zs for the impact parameter of 2.2~fm, where the last QF events are observed.
The light fragment mass and charge stay flat around $A_\mathrm{L}=101.5-107.1$ and $Z_\mathrm{L}=40.6-42.2$ until
a sudden drop for the largest impact parameter. Similar range of mass and charge combinations are also seen
in Fig.~\ref{fig:40Ca_stack_1} for a large energy range.
The root of this behavior may be due the fact that the Zr isotopes in the mass range 100$-$112 are strongly bound
with a large prolate deformation around $\beta_2=0.42$\cite{moller1995,lalazissis1999,blazkiewicz2005,hwang2006}.
Due to shell effects, these configurations may be energetically favorable during the QF dynamics.
However, since TDHF theory does not include fluctuations experimentally a distribution of masses will
be observed.

We have repeated the same set of TDHF calculations for the neutron-rich system
$^{48}$Ca+$^{238}$U, with the purpose of comparing the two systems.
In Figures~\ref{fig:48Ca_stack_1}a and ~\ref{fig:48Ca_stack_1}b we display the contact time
and the light fragment mass / charge as a function of the
center-of-mass energy for central collisions. These results are dramatically different
as compared to the $^{40}$Ca+$^{238}$U system: the quasifission region,
as evidenced by long contact time and large mass transfer, is confined
to a very narrow center-of-mass energy window, $E_{\mathrm{c.m.}}=202-203.5$~MeV.
In this QF energy region, the contact time varies between
(12.1$-$19.7)~zs.
Similarly, we find a light fragment mass range of $A_\mathrm{L}=78.7-102.4$
and charge range of $Z_\mathrm{L}=32.1-40.8$.
Using our microscopic approach we have also calculated the excitation energy of the
emerging fragments for the
neutron-rich system $^{48}$Ca+$^{238}$U as shown in Fig.~\ref{fig:48Ca_stack_1}c.
We find excitation energies
in the QF region up to $40$~MeV for the heavy fragment as compared to
$60$~MeV in the $^{40}$Ca+$^{238}$U reaction.
\begin{figure}[!htb]
	\centering
	\includegraphics*[width=8.6cm]{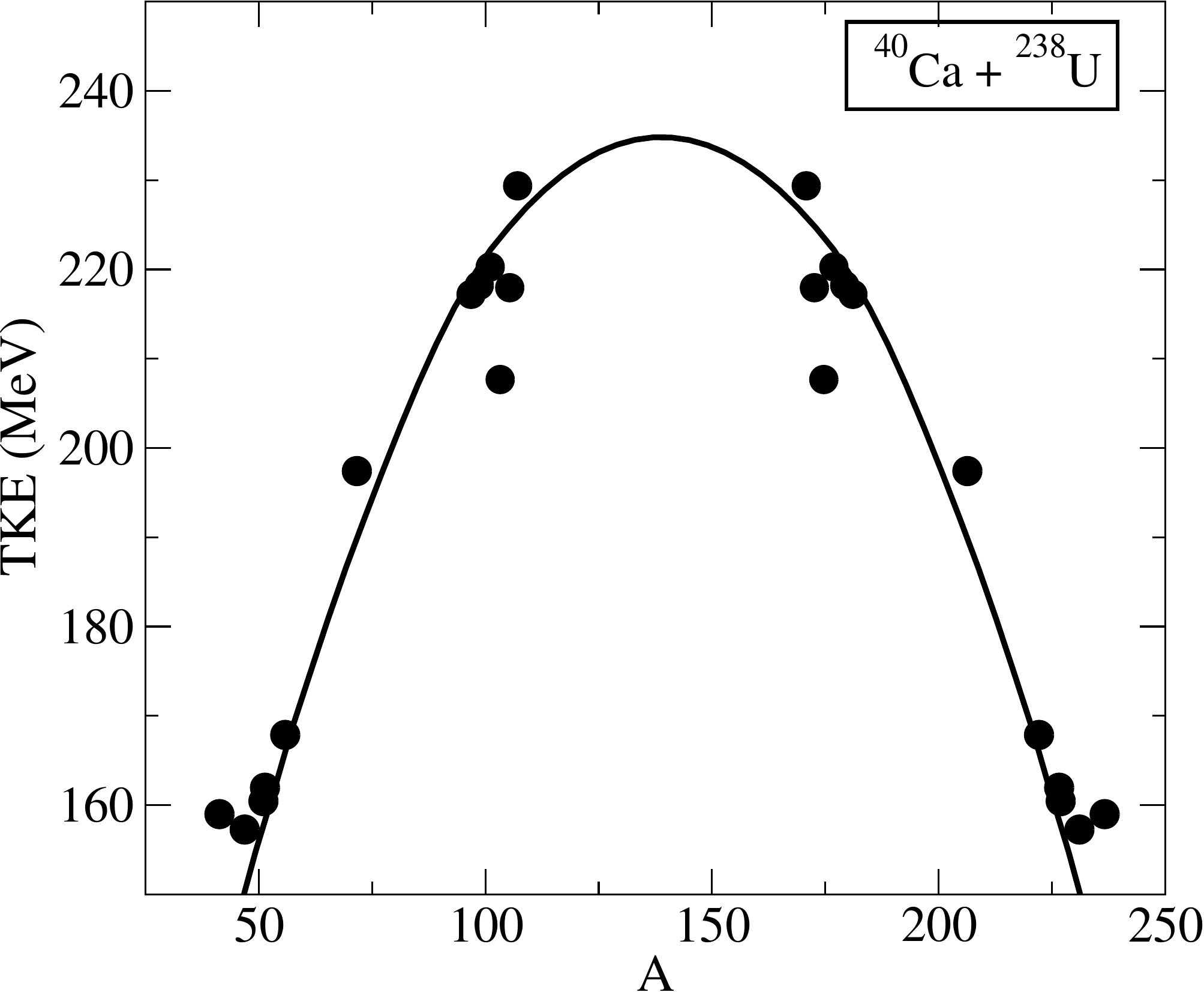}
	\caption{\protect TKE of both the light and heavy fragments formed in $^{40}$Ca+$^{238}$U central collisions.
		The filled circles represent results from TDHF calculations, and the solid line represents
		TKE values based on the Viola formula~\cite{viola1985}.}
	\label{fig:TKE}
\end{figure}

The contact times discussed above are long enough to enable the conversion of the initial relative kinetic
energy into internal excitations.
The total kinetic energy (TKE) distribution of the reaction products is one of the indicators of the
source of the observed fragments. For quasifission, the TKE distribution is expected to be
described by the Viola systematics~\cite{toke1985,nishio2012}.
This indicates the TKE's of final fragments are primarily due to their Coulomb repulsion and do not carry
a substantial portion of the initial relative kinetic energy.
Experimentally, the measured total kinetic energy of the quasifission fragments
in  $^{40,48}$Ca+$^{238}$U reactions is in relatively good agreement with the Viola systematics.
The TDHF approach contains one-body dissipation mechanisms which are dominant at near-barrier energies and can
be used to predict the final TKE of the fragments.
The TKE of the fragments formed in $^{40}$Ca+$^{238}$U have been computed for a range of central collisions up to $10\%$ above the barrier.
Figure~\ref{fig:TKE} shows that the TDHF predictions of TKE are in excellent agreement with the Viola systematics.
This is a further confirmation that the TDHF dynamics is providing a plausible description of the quasifission
process.

An important observation of the above results is that the neutron-rich $^{48}$Ca+$^{238}$U
system shows considerably less QF in comparison to the stable $^{40}$Ca+$^{238}$U system.
Similarly, the excitation energies of the emerging QF fragments have considerably
less intrinsic excitation. These results point to the conclusion that the neutron-rich
system has a higher probability for leading to the formation of a superheavy element, as it was
discovered experimentally.

\begin{figure}[!htb]
		\includegraphics*[scale=0.30]{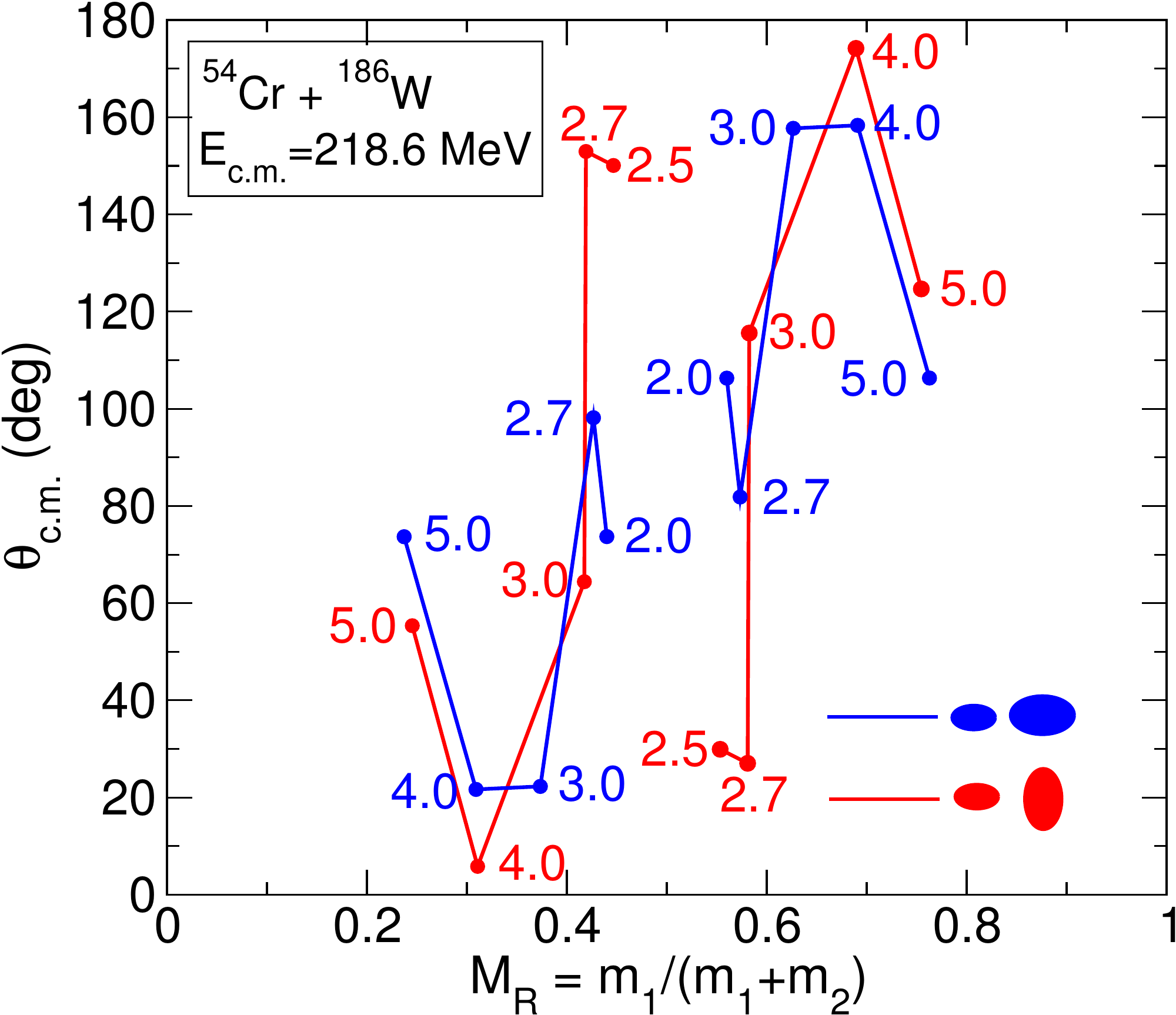}\hspace{0.1in}
		\includegraphics*[scale=0.36]{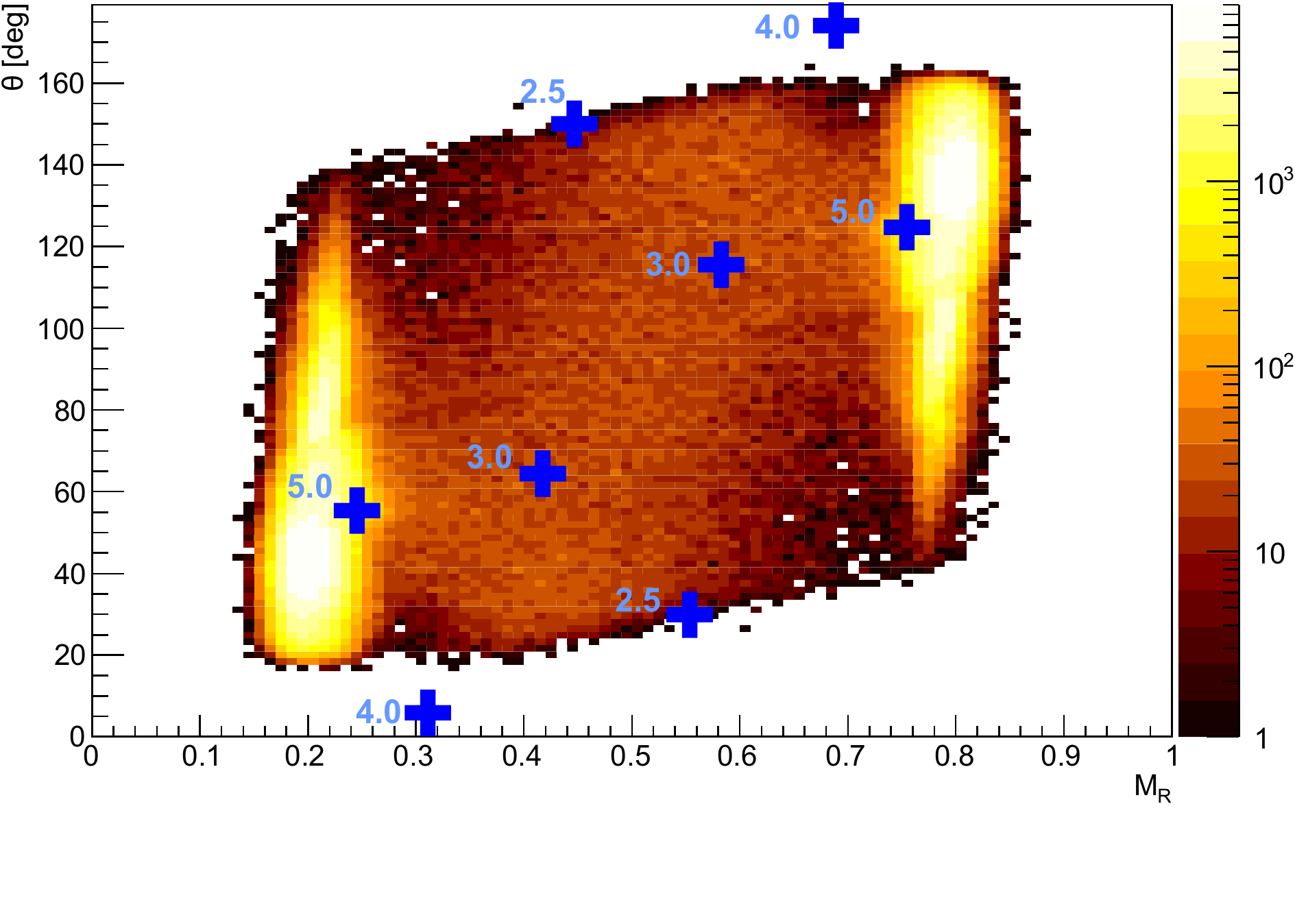}
	\caption{\protect Left: TDHF calculation of MAD for the quasifission products of the
		reaction $^{54}$Cr$ + ^{186}$W at $E_\mathrm{c.m.}=218.6$~MeV.
		Right: Experimental MADs~\cite{hammerton2014} corresponding to the same reaction. The blue crosses show
		the TDHF results for the tip-side collision with the corresponding impact parameter
		values in fm's.}
	\label{fig:mad}
\end{figure}

Another observable that can be studied using TDHF is the mass-angle distribution (MAD) for a
quasifission reaction. Experimental MAD's show the yield of mass-ratio, $M_R=m_1/(m_1+m_2)$,
as a function of the c.m. angle of the quasifission products with masses $m_1$ and $m_2$.
In the left plot of Fig.~\ref{fig:mad} we show the TDHF calculations of quasifission MADs for
the reaction $^{54}$Cr$ + ^{186}$W at $E_\mathrm{c.m.}=218.6$~MeV, corresponding to the two
orientations of the $^{186}$W nucleus. In the plot shown in the right side of Fig.~\ref{fig:mad}
corresponding experimental MADs are shown~\cite{hammerton2014}. The regions of MAD's near $M_R=0.2$ and $M_R=0.8$
correspond to elastic and quasielastic reactions, followed by transition to deep-inelastic
reactions and subsequently quasifission. The transition from deep-inelastic to quasifission
is correctly reproduced by TDHF as well as the general behavior of the MADs. Due to the fact
that TDHF is a deterministic theory it will only give us the most probable outcome or path
for the  MADs rather than a full distribution. Similar MAD's have been obtained for the
$^{40}$Ca+$^{238}$U system in Ref.~\cite{wakhle2014}.

\section{Ingredients for evaluating $P_{\mathrm{CN}}$}

In this section we will discuss the possibility of calculating some of the ingredients
that go into the computation of $P_{\mathrm{CN}}$ which is the probability
that the system evolves into a fused system rather than quasifission.
The two main references used in the discussion of $P_{\mathrm{CN}}$ for
quasifission are~\cite{tsang1983,yanez2013}.

\subsection{Moment of inertia}
During the collision process the nuclear densities, as described by TDHF time-evolution, undergo
complicated shape changes, rotations, etc. finally leading to two separated final
fragments identified as quasifission due to the long contact-time for the reaction
as well as the mass/charge of the fragments.

We have realized that the proper way to calculate the moment-of-inertia for such
time-dependent densities is to directly diagonalize the moment-of-inertia
tensor
\begin{equation}
\Im_{ij}/m = \int~d^3r\;\rho_{TDHF}(\mathbf{r},t) (r^2\delta_{ij}-x_ix_j)\;,
\end{equation}
where $\rho_{TDHF}$ is the number-density in units of $(N/fm^3)$, $m$ is the
nucleon mass, and $x_i$ denote the Cartesian coordinates. The TDHF calculations
are done in three-dimensional Cartesian geometry~\cite{umar2006c}.
Numerical diagonalization of this $3\times 3$ matrix gives the $3$ eigenvalues,
one corresponding the the moment-of-inertia for the nuclear system rotating
about the symmetry axis, and the other two for rotations about axes perpendicular
to the symmetry axis. We denote these by $\Im_{\parallel}$ and $\Im_{\perp}$.
Naturally, for triaxial density distributions the two perpendicular components
are not exactly equal but for practical calculations they are close enough
and always larger than the parallel component.

Using the time-dependent moment-of-inertial obtained from the TDHF collision
one can calculate the so-called effective moment-of-inertia
\begin{equation}
\frac{1}{\Im_{eff}}=\frac{1}{\Im_{\parallel}}-\frac{1}{\Im_{\perp}}\;.
\end{equation}
In literature~\cite{tsang1983,yanez2013} what is usually given is the ratio
$\Im_0/\Im_{eff}$ at the saddle point of the fission barrier, where $\Im_0$ is the
moment-of-inertia of spherical nucleus with the same mass. This ratio is to be
constant for impact parameters leading to quasifission $(J>J_{CN})$, where
$J_{CN}$ is the largest $J$ value resulting in compound nucleus formation.
The expression for the moment-of-inertia for a rigid sphere is given by
$\Im_0/m=2/5AR_0^2$, which in units of $\hbar^2 MeV^{-1}$ can also be written as
\begin{equation}
\Im_0=\hbar^2 (2/5AR_0^2)/(\hbar^2/m)
\end{equation}
In Ref.~\cite{tsang1983} the $R_0$ was chosen to be $R_0=1.225A^{1/3}$. With the
choice of $\hbar^2/m=41.471$~MeV$\cdot$fm$^2$, corresponding to the value used
in the Skyrme SLy4d interaction, for $A=286$ we get $\Im_0=179.693$~$\hbar^2\cdot$MeV$^{-1}$

\begin{figure}[!htb]
        \centering
	\includegraphics*[width=8.6cm]{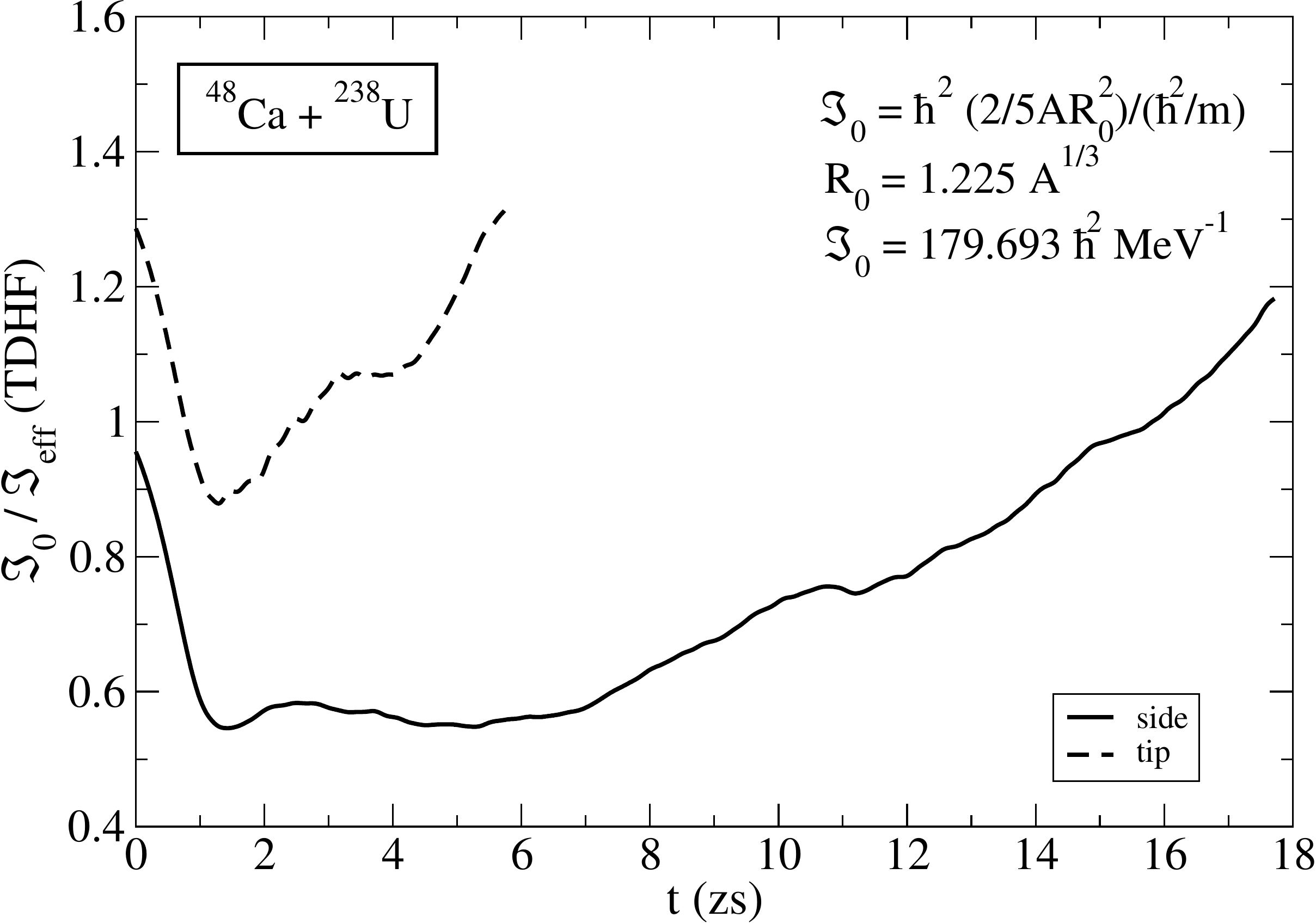}
	\caption{\protect
		TDHF results showing the time-dependence of the ratio $\Im_0/\Im_{eff}$ for the $^{48}$Ca~+~$^{238}$U system
		at energy $E_\mathrm{c.m.}=203$~MeV and zero impact parameter.
	}
	\label{fig:iratio}
\end{figure}

In Fig.~\ref{fig:iratio} we show the time-evolution of the ratio $\Im_0/\Im_{eff}$
for the TDHF collision of the $^{48}$Ca~+~$^{238}$U system
for central collision at an energy $E_\mathrm{c.m.}=203$~MeV.
We see that during the times that possibly correspond to the vicinity of the
saddle point the ratio appears to be smaller than the traditionally used value of $1.5$.
We will come back to the discussion of finding the saddle point later in this document.
Before we end this section we should also point out that we having the numerical values
for $\Im_{\perp}$ also allows the computation of the rotational energy
\begin{equation}
E_{rot}=\frac{\ell(\ell+1)\hbar^2}{2\Im_{\perp}}\;.
\end{equation}

\subsection{Quasifission barrier}
Recently, the newly developed density-constrained TDHF method has proven to be a
powerful method of obtaining fusion barriers microscopically from TDHF time-evolution
of the nuclear densities. This is a parameter free way of obtaining these barriers.

In principle, the same approach can be used to compute the underlying barrier during
the quasifission dynamics. From this we may be able to get most of the other
ingredients of computing
\begin{equation}
K_{0,\ell}=T\Im_{eff}/\hbar^2\;,
\end{equation}
where $T$ is the nuclear temperature involving various quantities such as $E^*$, barrier height,
and others. In practice the $\ell$ dependence of this expression is ignored, which may be
a reasonable approximation.
The computation of the barrier will be very time-consuming but it may give us a better
understanding of the quasifission process.

\section{Conclusions}

Recent TDHF calculations of phenomena related to SHE searches show that TDHF can be
a valuable tool for elucidating some of the underlying physics problems encountered.
As a microscopic theory with no free parameters, where the effective nucleon-nucleon
interaction is only fitted to the static properties of a few nuclei, these results
are very promising.

\section*{Acknowledgments}
We gratefully acknowledge discussions with C. Simenel, Z. Kohley, and W. Loveland.
This work has been supported by the U.S. Department of Energy under Grant No.
DE-FG02-96ER40975 with Vanderbilt University.


\bibliography{np_she.bib}

\end{document}